\documentclass[twocolumn,pre,showpacs,amsmath,amssymb]{revtex4-1}
\usepackage{units}
\usepackage{graphicx}
\newcommand{\be}{\begin{eqnarray}}
\newcommand{\ee}{\end{eqnarray}}

\newcommand{\veps}{\varepsilon}
\newcommand{\bfp}{\mathbf p}
\newcommand{\bfq}{\mathbf q}
\begin{document}

\title{Evolution and Stability of Altruist Strategies in Microbial Games}
\author{Christoph Adami}
\email{adami@msu.edu}
\affiliation{Department of Microbiology \& Molecular Genetics}
\affiliation{Department of Physics \& Astronomy}
\affiliation{BEACON Centre for the Study of Evolution in Action\\
Michigan State University, East Lansing, MI 48824}
\affiliation{Keck Graduate Institute, Claremont, CA 91711}
\author{Jory Schossau}
\affiliation{Department of Computer Science \& Engineering}
\affiliation{BEACON Centre for the Study of Evolution in Action\\
Michigan State University, East Lansing, MI 48824}
\author{Arend Hintze}
\affiliation{Department of Computer Science \& Engineering}
\affiliation{BEACON Centre for the Study of Evolution in Action\\
Michigan State University, East Lansing, MI 48824}



\begin{abstract}When microbes compete for limited resources, they often engage in chemical warfare using bacterial toxins. This competition can be understood in terms of evolutionary game theory (EGT). We study the predictions of EGT for the bacterial ``suicide bomber" game  in terms of the phase portraits of population dynamics, for parameter combinations that cover all interesting games for two-players, and seven of the 38 possible phase portraits of the three-player game. We compare these predictions to simulations of these competitions in finite well-mixed populations, but also allowing for probabilistic rather than pure strategies, as well as Darwinian adaptation over tens of thousands of generations.  We find that Darwinian evolution of probabilistic strategies stabilizes games of the rock-paper-scissors type that emerge for parameters describing realistic bacterial populations, and point to ways in which the population fixed point can be selected by changing those parameters.  
\end{abstract}
\pacs{87.23.Kg,87.10.Mn,87.18.Hf}
\maketitle
\section{Introduction}Evolutionary Game Theory (EGT)~\cite{MaynardSmith1982,HofbauerSigmund1998,Nowak2006} has become one of the pillars of evolutionary biology because it is a mathematically accessible framework that can account for the strategic aspect of frequency-dependent fitness, that is, if the fitness of a genotype depends on the frequency of other genotypes in the population. EGT is particularly useful when dealing with populations that include a mix of different strategies, and takes into account the concept of ``inclusive fitness" that encompasses how an organism contributes to the fitness of other genetically similar or even different types~\cite{Hamilton1964,Grafen1985,Queller1992}. Consequently, EGT has been used to study the conditions for the emergence, maintenance, and evolution of cooperation as well as altruism~\cite{FletcherDoebeli2006,FletcherZwick2006,FletcherDoebeli2009}. Of particular importance is the application of EGT to microbial communities (see, e.g.,~\cite{Frey2010}). The competitive interaction between microbes (and even viruses~\cite{TurnerChao1999,TurnerChao2003}) is often best described within the language of games~\cite{GreigTravisano2004,Goreetal2009,Chuangetal2009}, and they display cooperative and other types of social behavior~\cite{VelicerVos2009}, cheating~\cite{Strassmannetal2000}, and even an extreme form of altruism where some community members sacrifice themselves for the sake of others. Another hallmark of microbial community dynamics is the observation of cyclically competing species with non-transitive relationships~\cite{Kerretal2002}.
Of particular interest is EGT's prediction of evolutionary fixed points for some dynamical interactions but not others, and characterization of the stability of orbits within the phase space of strategies. However, how these predictions compare to realistic dynamics of adapting populations of cells, in particular those making stochastic decisions~\cite{Perkinsetal2009}, is not always clear. Here, we study the predictions of EGT as applied to a simple bacterial system where different strategies of survival potentially coexist in the same population, or where strategic decisions are made stochastically rather than deterministically.

Among the weapons that bacteria use to battle each other is the production of {\em colicin}, a bacterial toxin that is lethal to most strains of {\it E. coli} bacteria. Those strains that can produce the toxin are usually unharmed by it due to a resistance gene encoded on a plasmid within the cell. The individual that actually deploys the toxin, however, pays the ultimate price as that cell literally explodes to distribute the toxin to as large a fraction of the population as possible. This ultimate act of altruism is beneficial to the ``suicide bomber"'s kin because it frees up the resource that both sensitive and resistant types are using, for the sole benefit of the resistant types. 
This system has been studied experimentally~\cite{ChaoLevin1981,Kerretal2002}, and its dynamics studied in terms of payoff matrices~\cite{LenskiVelicer2000}. Some of the dynamics we observe falls into the ``Rock-Paper-Scissors" (RPS) category of games, which have been studied analytically and experimentally~\cite{MaynardSmith1982,HofbauerSigmund1998,Kerretal2002,SzaboHauert2002,SzolnokiSzabo2004,NeumannSchuster2007,ClaussenTraulsen2008,Frey2010} (see also the review~\cite{SzaboFath2007}). We study the EGT predictions of the equilibrium frequency of strategies in the population (when this equilibrium exists), how this equilibrium is modified when additional strategies can be produced via mutations, and how EGT predictions fare when population sizes are finite. To study these predictions we use different numerical simulation methods, and in particular study strategies that evolve via mutation and selection in a purely Darwinian manner~\cite{Iliopoulosetal2010}. Furthermore, because the expression of the toxin is probabilistic in nature (experimentally, only between 1\% and 5\% of the bacteria that carry the toxin plasmid actually express it~\cite{ChaoLevin1981,LenskiVelicer2000}), we study fully stochastic strategies with evolvable probabilities. In the following, we first introduce the notation and a discussion of equilibrium points in the well-known two-player game, setting the stage for a similar analysis of the three-player game.

\section{Two-Player Suicide Bomber Game}


Evolutionary game theory makes accurate predictions about the outcome of two-player games, whether the strategies are deterministic (`pure' strategies) or probabilistic (`mixed' strategies, where a player uses different pure strategies with different probabilities).  The central concept of EGT is the ``evolutionary stable strategy" (ESS): If only one  strategy is determined to be an ESS, we should find this and only this strategy to be the winner in a competition. In a game with two strategies $I$ and $J$, $I$ is an ESS  if the payoff $E(I,I)$ when playing itself is larger than the payoff $E(J,I)$ between any other strategy $J$ and $I$, i.e.,~\cite{MaynardSmith1982}
\be 
I\ {\rm is}\  {\rm ESS}\  {\rm if}\  E(I,I)>E(J,I)\;. \label{ess1}
\ee
In case $E(I,I)=E(J,I)$, then $I$ is an ESS if the strategy plays better against any other strategy $J$ then that other strategy fares against itself:
\be
E(I,J)>E(J,J)\ {\rm when}\ E(I,I)=E(J,I). \label{ess2}
 \ee
 In principle, a mixture of strategies can be an ESS. So introduce the population mixture described by the vector $\bfp=(p_I,p_J)$ where $p_I$ is the population fraction of strategy $I$ and $p_I+p_J=1$.  Then $\hat \bfp $ is an ESS if (and only if) for all $\bfq$ (see e.g., ~\cite{HofbauerSigmund1998})
  \be
 \hat \bfp \cdot E \hat \bfp \geq \bfq\cdot E\hat \bfp \label{Nash}
 \;,
 \ee 
and at the same time
\be 
{\rm if} \  \bfq\neq \hat\bfp\ {\rm  and}\ \hat \bfp \cdot E \hat \bfp =  \bfq\cdot E\hat \bfp  ,\ 
{\rm then}\  \hat \bfp \cdot E \bfq > \bfq\cdot E \bfq\;. \ \ \ \ \ \  \label{equi}
\ee
Here, $E$ is the payoff matrix introduced above. Eqs.~(\ref{Nash},\ref{equi}) are just the population generalizations of (\ref{ess1},\ref{ess2}). Condition (\ref{Nash}) defines the Nash equilibrium point of the population, while (\ref{equi}) is the stability condition for the Nash equilibrium. It turns out, however, that the concept of an ESS is not as general as one might wish, because while every ESS is a stable attractor of the population dynamics (in the language of dynamical systems theory), not every stable attractor is an ESS~\cite{TaylorJonker1978,Hofbaueretal1979,Zeeman1980}. Thus, instead of focusing on ESSs, in the following we study the fixed points and phase portraits of the dynamics that the payoff matrix induces.

In the two-player ``suicide bomber" game of colicin warfare, the payoff between the wild-type (which we denote here as strain `00') and the colicin-producing (but resistant) type `RT' is such that $E(00,00)=1$, but $E(00,{\rm RT})=0$, that is, a wildtype strain is killed by a strain expressing the toxin. On the other hand, RT is itself inferior to the wildtype because it carries the cost of that resistance, so $E({\rm RT},{\rm RT})<1$. Finally, $E({\rm RT},00)>1$, expressing the advantage the colicin producer has due to the suicidal behavior of its kin~\cite{LenskiVelicer2000}. Note that the cost of resistance is typically of the order of 15\% of wild-type fitness, but can be as small as 1\%~\cite{FeldgardenRiley1999} or as large as 60\%~\cite{Nahumetal2011}. The cost of producing the colicin (including the reduction of growth rate by cell lysing) may be even higher, depending on the frequency with which colicin is being produced~\cite{Nahumetal2011}. 
Our notation `00' and `RT' for the two different types comes from the observation that the production and the resistance of colicin in bacteria are usually encoded by two different genes (`R' for resistance, and `T' for toxin), but most often within the same plasmid. 

\begin{figure}[htbp] 
   \centering
   \includegraphics[width=8cm]{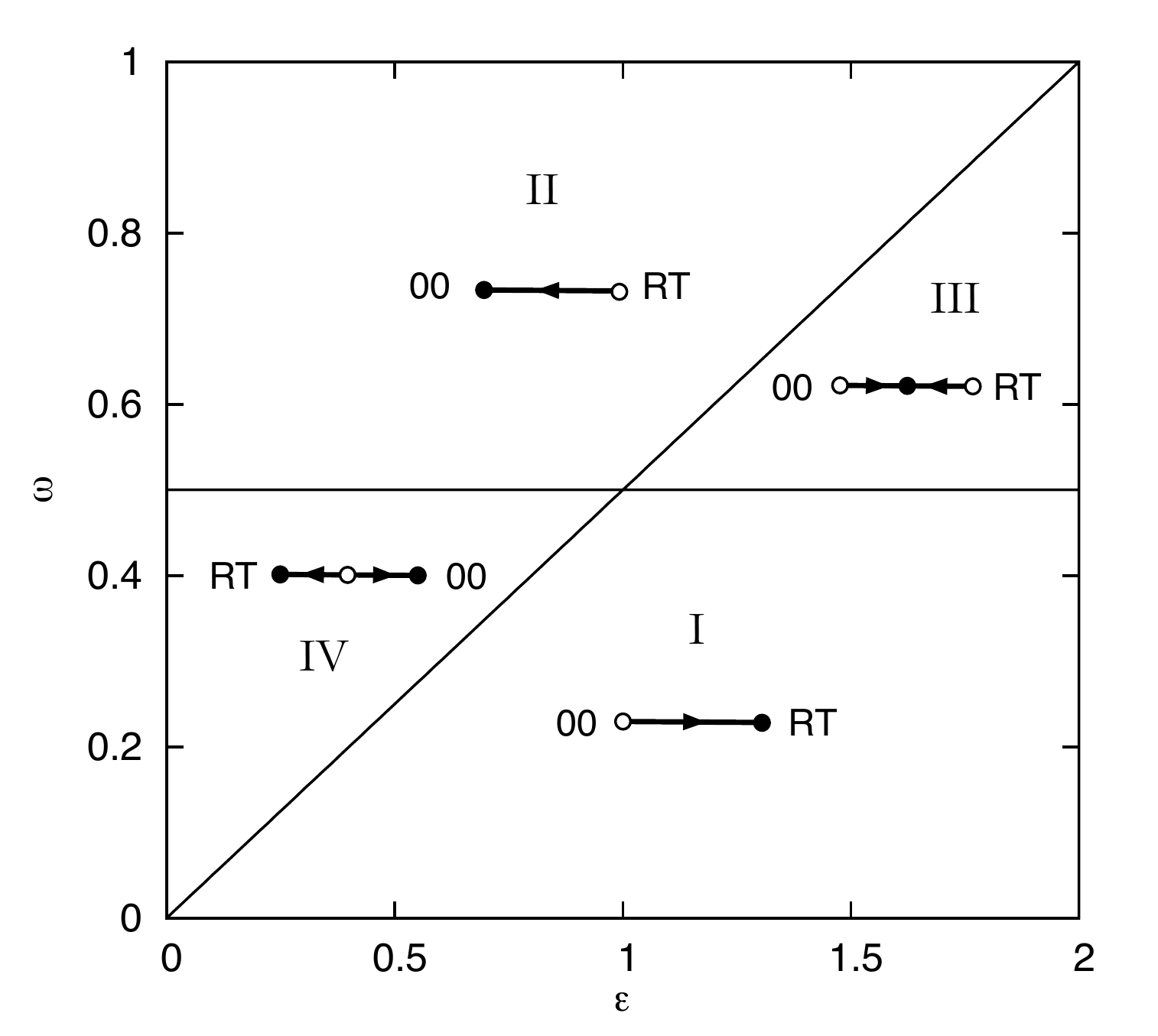} 
   \caption{Fixed points and flow for different two-player games as a function of game parameters, for $\omega\in (0,1)$ and $\veps\in(0,2)$. Region I: Prisoner's Dilemma, region II: ``Harmony" game, region III: Snowdrift game, region IV: ``coordination" game (see, e.g.,~\protect\cite{Frey2010}). }
   \label{2dyn}
\end{figure}

We study a model in which two parameters govern the interaction between types: the benefit $\varepsilon\geq0$ of expressing the toxin and the cost $\omega\geq0$, and vary these parameters systematically. In principle, the two genes R and T could each carry a different cost, but for most of the calculations in this study we will keep them the same. The ESS is determined by the following payoff matrix $E$:
\be
\bordermatrix{\mbox{} & 00& {\rm RT}  \cr
                            00   & 1  & 0 \cr
                            {\rm RT}     &1-2\omega+\veps &  1-2\omega }
                           \;. \label{Mat1}
\ee
The advantage $\veps$ is a variable that depends on the spatial structure of the environment and the distribution of types in it, because suicide behavior will be more effective when the resource that is freed up is more likely to be used by the resistant kind. Thus, we study what strategy is an ESS as a function of the parameters $\omega$ and $\veps$. 
In terms of the more conventional notation of two-player games we have $R=1$, $S=0$, $T=1-2\omega+\veps$, and $P= 1-2\omega$, that is, 00 is the cooperating type. Because payoff matrices that only differ when adding a constant to each column induce the same dynamics, we can bring the matrix into normal form so that the diagonal vanishes:

\be
  E= \left(\begin{matrix} 
      0 &2\omega-1  \\
      \veps-2\omega & 0 \\
   \end{matrix}\right)
                           \;. \label{Mat2}
\ee
The fixed points, stability, and phase portrait of such games has been solved (see, e.g,~\cite{Zeeman1980,SzaboFath2007}). 
The dynamics falls into four regions that harbor three different phase portraits. In Region I, defined by $\omega<1/2, \omega<\veps/2$ we find the standard prisoner's dilemma, where the ``defecting" strategy is ESS as well as an attractor. We indicate in Fig.~\ref{2dyn} the parameter region for this dynamics, along with a pictogram that describes the phase portrait with the convention that solid circles represent attractors and open circles denote repellers. Thus, in this language the wild-type is the cooperator and the suicide bomber the defector. For $\omega>1/2, \omega>\veps/2$ (region II) there is no dilemma and the wild-type is the ESS. When $\omega>1/2, \omega<\veps/2$ (region III), the game is an ``anti-coordination game" sometimes called ``snowdrift" or ``Hawk-Dove", giving rise to a stable population mixture of strategies as indicated by the fixed point along the trajectory connecting the wild-type and the suicidal type RT, with $p_{00}=\frac{2\omega-1}{\veps-1}$. 
If on the other hand $\omega<\frac12$ and $\omega>\veps/2$ (region IV), both strategies are an ESS as well as attractors, and the ultimate winner depends on which strategy is more abundant at the start of the competition, as was observed in the Chao-Levin experiment~\cite{ChaoLevin1981} in a well-mixed environment. Note that this regime corresponds to a ``coordination" game. If strategies are probabilistic, equations (\ref{Nash}) and (\ref{equi}) predict dominance of a mixed strategy with the probability to engage in `00' play given by the $p_{00}$ shown above~\cite{MaynardSmith1982}.   
The different predicted phases are consistent with those observed for competitions between sensitive and resistant types on a lattice~\cite{Iwasaetal1998}.
 \begin{figure}[htbp] 
    \centering
    \includegraphics[width=3.5in]{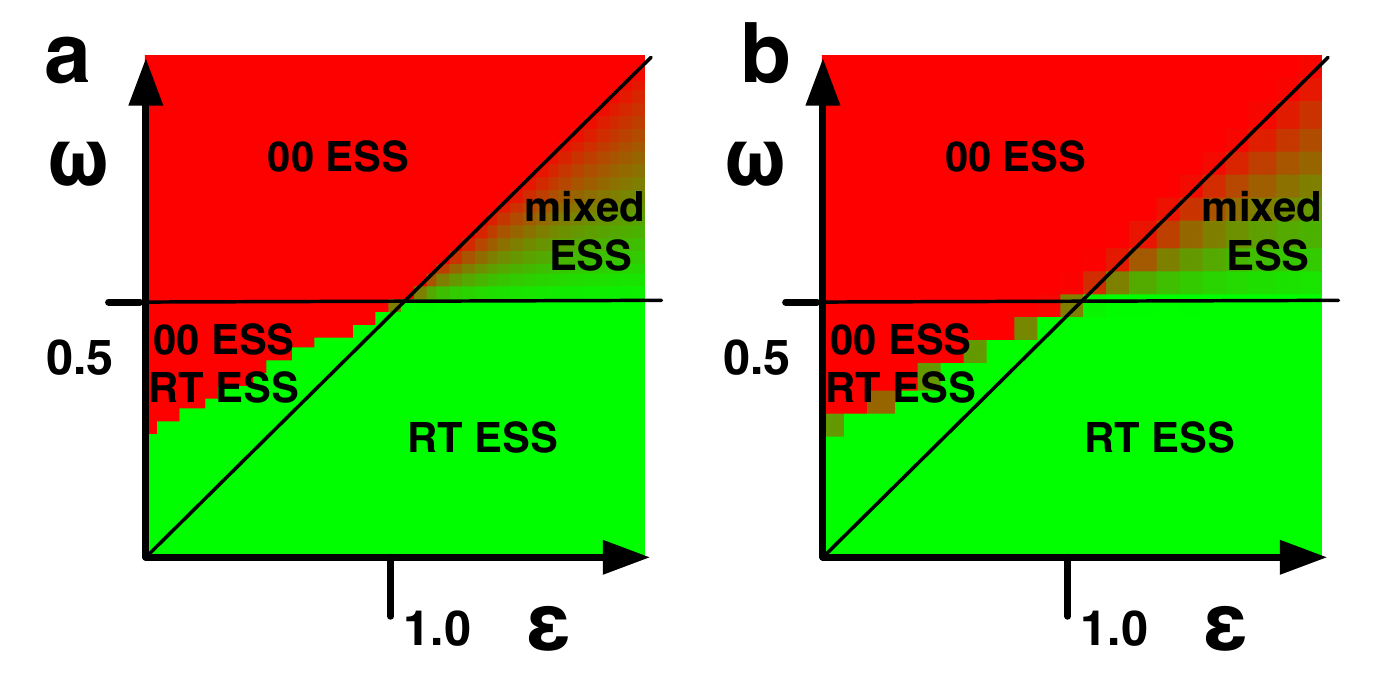} 
    \caption{ESS as a function of the parameters $\veps$ and $\omega$. (a): Numerical simulation of the population frequencies based on replicator equations, with both types at equal frequency initially. (b) Simulation of a finite well-mixed population of agents that encode the strategies 00 and RT deterministically, with the same initial conditions as in (a) [50 independent simulations for each of $21\times21$ different parameter combinations]. In (a) and (b), red and green intensity indicates the frequency of strategy 00 and RT in the population, respectively. The separation between red and green in region IV is due to the choice of initial condition, and does not represent a phase boundary.}
    \label{fig:fig1}
 \end{figure}
We can confirm the predictions of EGT by numerical simulation of the population dynamics using replicator equations~\cite{MaynardSmith1982,HofbauerSigmund1998}:
 \be
 \dot p_{00}(t)&=&p_{00}(t) (w_{00}-{\bar w})\;,\label{update1}\\
 \bar w &=& p_{00}(t)w_{00}+\left(1-p_{00}(t)\right)w_{\rm RT}\;, \label{update2}
 \ee
 with 
 \be
 w_{00}&=&p_{00}E(00,00)+(1-p_{00})E(00,{\rm RT})\;,\\
 w_{\rm RT}&=&p_{00}E({\rm RT},00)+(1-p_{00})E({\rm RT},{\rm RT})\;.
 \ee 
We show the result of this simulation in Fig.~\ref{fig:fig1}a,  where both strategies were initialized with $p=0.5$. The replicator equation simulations recapitulate the phase portraits in Fig.~\ref{2dyn}.

We can also compare the EGT prediction to a simulation of the evolution of agents that receive payoffs (\ref{Mat1}) in a finite (but large) well-mixed population of 16,384 interacting agents. Here, each player's strategy is determined by a `genome' with a single locus $p_{00}$ that stands for the probability to engage in action 00. For each agent, we randomly pick four opponents, and the aggregate score against them is used as a fitness. We replace 2\% of the population at each update, using a death-birth Moran process~\cite{Moran1962} (for more details on the simulations, see Section~\ref{stoch}). If we start the population with 50\% genotypes reflecting the wildtype ($p_{00}=1$) and 50\% expressing the RT phenotype ($p_{00}=0$) without any mutations so that the pure strategies compete, we recapitulate the theoretical results as well as the replicator equation simulations  (see Fig.~\ref{fig:fig1}b).  If strategies can be mixed and we allow mutations on $p_{00}$, we find $p_{00}\to\frac{2\omega-1}{\epsilon-1}$ as expected (data not shown). Indeed, it can be shown that for two-player games, the predicted population fraction fixed points are equal to the fixed points in the space of decision probabilities, and that the stability of these fixed points also coincides~\cite{MaynardSmith1982}. For the three-player game, the fixed points of deterministic and stochastic strategic also coincide, but the stability conditions do not.  

\section{Three-Player Game}
So far we studied the adaptive change of genotype frequencies and expression probabilities, but we have not addressed the consequences of macroscopic mutations. While the toxin gene and the resistance gene are usually tightly linked on the same plasmid~\cite{Jamesetal1996}, cells can acquire resistance to the toxin without carrying the plasmid, for example by changes to a receptor or the membrane protein that imports the colicin. However, such changes are usually costly because the same proteins are also involved in importing nutrients into the cell. 

To take mutations that create new strategies into account, we introduce the additional type `R0' (the type `0T', which we simulate using Darwinian evolution below, does not play a role here because it is never an ESS), and study how the possibility of mutating into this type affects the stability of the fixed points. In a general payoff matrix for the suicide-bomber game, non-resistant cells suffer an effect $\Delta\geq0$ from exposure to the colicin, and carrying the toxin gene incurs a cost $\omega$ while resistance decreases fitness by $\delta\geq0$. As before, the advantage of the RT phenotype is $\veps$. The payoff $E$ is now:
\be
\bordermatrix{\mbox{} & 00& {\rm R0} & {\rm RT}  \cr
                            00     & 1  & 1  & 1-\Delta \cr
                            {\rm R0}     &1-\delta & 1-\delta &1-\delta\cr
                            {\rm RT}    & 1-\delta-\omega +\veps&  1-\delta-\omega &1-\delta-\omega}
                           \;. \label{Mat3}
\ee
With these values, the resistant type R0 is superior to the resistant toxin producer RT but inferior to the wildtype 00 because of the cost of resistance. The dynamics of these three strains can become non-transitive so that all three strain can outcompete each other in a classic  rock-paper-scissors (RPS) dynamic~\cite{Kerretal2002}.  We can study the fixed points and phase portraits of this game theoretically, as well as via agent-based simulations. For simplicity, we take $\Delta=1$ as before, and restrict ourselves to $\delta=\omega$ (cost of resistance equal to cost of toxin). 
Furthermore, we can normalize the payoffs such that the diagonal vanishes, so that the payoff  matrix becomes
\be
E=\left(   \begin{matrix} 
      0 & \omega & 2\omega-1 \\
      -\omega & 0& \omega \\
      \veps-2\omega& -\omega& 0
   \end{matrix}\right)\;. \label{3play}
   \ee
\subsection{Stability and Zeeman classes}
\begin{figure}[htbp] 
   \centering
   \includegraphics[width=8cm]{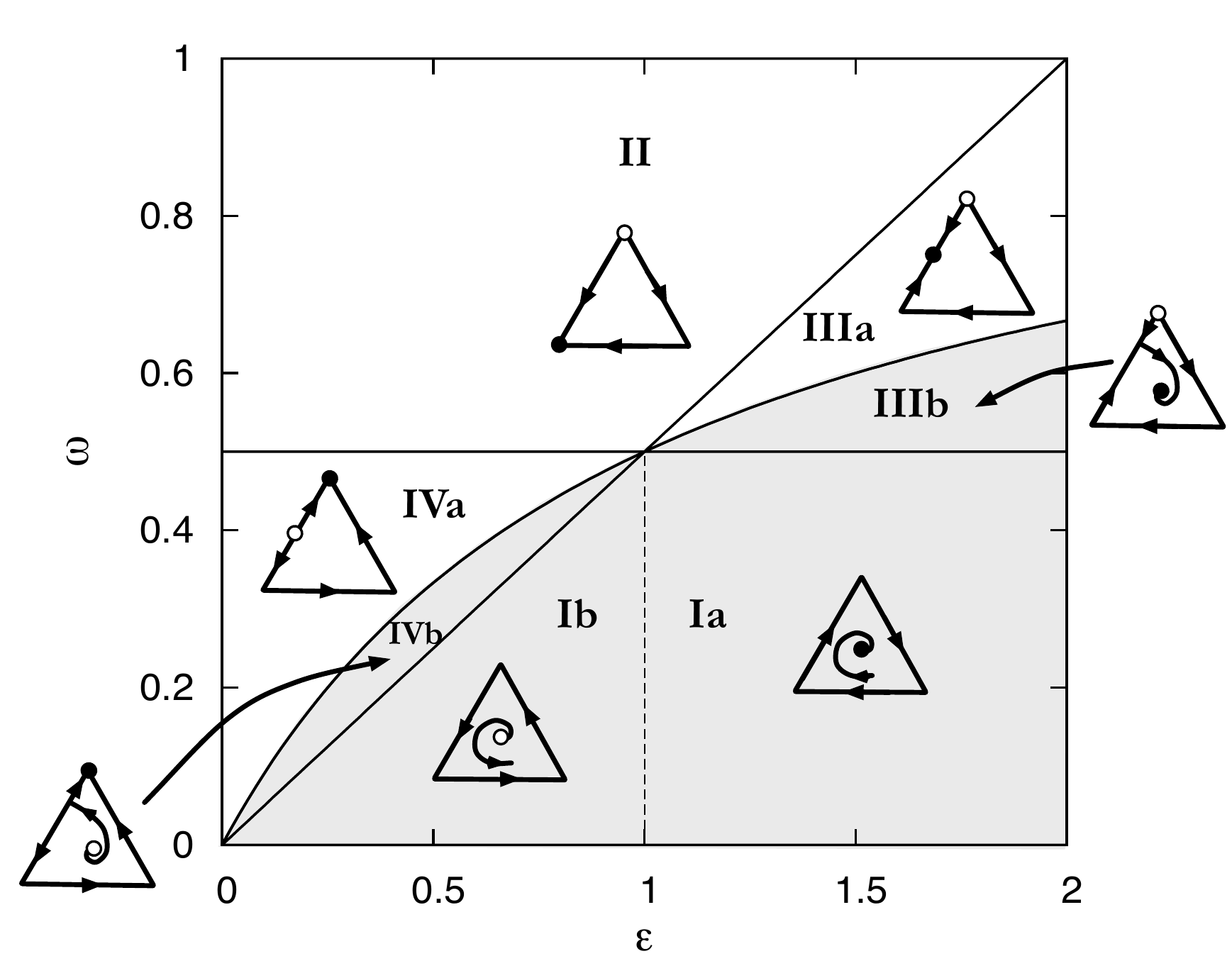} 
   \caption{Phase portraits of stable dynamics for the three-player game using payoffs (\ref{3play}). The shaded parameter region $\omega<{\veps}/({\veps+1})$ has an interior fixed point that can be repulsive or attractive. In Zeeman's phase portrait pictograms~\cite{Zeeman1980}, arrows denote the flow on the boundary of the simplex, solid circles are attractors and open circles are repellers. All fixed points on the boundary and the interior are indicated. The Zeeman classes for each region are indicated in Table~\ref{tab:zeeman}.  }
   \label{zeeman}
\end{figure}

The population dynamics of three-player games has been solved completely~\cite{Zeeman1980}, and crucially depends on the structure of fixed points in the interior or on the boundary of the 2-simplex $\Delta$ defined by the probabilities $(p_{00},p_{\rm R0},p_{\rm RT})$ with the constraint $p_{00}+p_{\rm R0}+p_{\rm RT}=1$. Using this 2-simplex, a simplified phase portrait of the dynamics can be constructed using a notation depicting attractors and repellers just as in Fig.~\ref{2dyn}. Zeeman showed that the dynamics fall into ten combinatorial classes (up to sign-reversal of each element of the payoff matrix),  depending on the number of fixed points in the interior and on the boundary of the simplex. Each combinatorial class itself may contain one or more stable classes giving rise to 19 different ``stable" phase portraits plus their ``inverses", where all flow directions are reversed and attractors are replaced by repellers and vice versa. Here, ``stable" means that the phase portrait does not change drastically within a neighborhood of the parameter values defining the game.~\cite{Zeeman1980}. The three-player suicide-bomber game as defined by (\ref{3play}) displays seven of these classes in seven regions, as shown in Fig.~\ref{zeeman} and listed in Table~\ref{tab:zeeman}. 
An interior fixed point 
only exists for the parameter region $\omega\leq\frac{\veps}{\veps+1}$,  shown shaded in Fig.~\ref{zeeman}, so that regions III and IV of the two-player game now contain dynamics with a fixed point, and dynamics without.   In region II, the wild-type 00 is a stable attractor just as in the two-player game. The only difference is that there are now many different paths from strategy RT to 00 that include R0 as an intermediate type. Region I (corresponding to the Prisoner's Dilemma region in the two-player game) has an interior fixed point (the Nash equilibrium) for payoff matrix (\ref{3play})\footnote{For the general payoff matrix~(\ref{Mat3}), the fixed point is $p_{00}=\frac{\delta}\veps, p_{\rm R0}=1-\frac\delta\veps-\frac{\omega}\Delta, p_{\rm RT}=\frac{\omega}\Delta$.}
\be
p_{00}&=& \frac{\omega}\veps\;,\label{p1}\\ 
p_{\rm R0}&=&1-\frac\omega\veps-\omega\;,\label{p2}\\
p_{\rm RT}&=&\omega\;. \label{p3}
\ee
that is attractive when $\det E>0$ (region Ia) and repulsive for $\det E<0$~\cite{Zeeman1980} (region Ib). For our payoff matrix, the boundary is given by the line $\veps=1$, as indicated in Fig.~\ref{zeeman}. The dynamic in this region is a rock-paper-scissors (RPS) game that is stable if $\veps>1$, that is, the population fractions approach the interior fixed point given by Eqs.~(\ref{p1}-\ref{p3}). For $\veps<1$ the RPS game is unstable, and we expect to observe heteroclinic orbits that ultimately pass through the single-strategy fixed points. These oscillations are similar to those observed by May and Leonard in the Lotka-Volterra dynamics of three species~\cite{MayLeonard1975}. 

\begin{table}[htbp]
   \centering    
   \begin{tabular}{|c|c|} 
   \hline
                Region   & Zeeman class  \\
   \hline
         Ia     &  1 \cr
         Ib &  -1\cr
         II & 2 \cr
         IIIa & $5_2$\cr
         IIIb & $5_1$\cr
         IVa & $-5_2$\cr
         IVb & $-5_1$ \cr
         \hline
   \end{tabular}
   \caption{Zeeman classes~\cite{Zeeman1980} for the seven regions with stable dynamics. Note  that regions IIIa and IVa, as well as IIIb and IVb are sign-opposites, that is, the matrix of signs of entries in the normalized payoff matrix (up to permutations) is reversed.}
   \label{tab:zeeman}
\end{table}
The ``snowdrift" region of the two-player game is now divided into region IIIa with a stable fixed point that is a mixture of RT and 00 only as in the two-player game, and a region IIIb with a stable interior fixed point given by Eqs.~(\ref{p1}-\ref{p3}). The anti-coordination region of the two-player game (regions IVa and IVb off the three-player game) show just the inverse of the dynamics in regions IIIa and IIIb, as outlined in Figure~\ref{zeeman}. Thus, in region IVb the R0-type is short-lived and the game reverts to a two-player game with its attendant stability properties (just as in region IIIa).

A simulation of strategy competition using the replicator equations validates the phase portraits of the various Zeeman classes, as seen in Fig.~\ref{fig4}a. For these simulations, we used starting conditions where all three strategies are equiprobable, and stopped after a finite number of updates of the equations. As a consequence, a particular strategy appears to be dominant for the unstable RPS game (blue in Fig.~\ref{fig4}a, left) even though in fact the trajectory cycles through the three pure strategies. The expanding orbits of parameter combination A in region Ia become shrinking orbits for region Ib (see parameter combination B), but because the orbits still touch the boundary, there is a chance of strategy extinction in finite populations. Parameter combination C in Fig.~\ref{fig4}a is also in region Ib, but because the orbits are tighter, extinction is unlikely even in finite populations. If $\veps=1$ in region I, the orbits are limit cycles encircling the fixed points, but this dynamic is not ``stable" in the sense of Zeeman as infinitesimal changes in the payoffs will change the trajectories qualitatively. Indeed, Zeeman proved that there are no structurally stable limit cycles in three-player games~\cite{Zeeman1980}. As $\veps$ passes through the critical value $\veps=1$, the flow exhibits a Hopf bifurcation~\cite{MarsdenMcCracken1976}. 

Parameter combination D in Fig.~\ref{fig4}a lies in region IIIb which has an interior fixed point. This point lies close to the boundary of region IIIA in which the fixed point is on the edge (complete extinction of strategy R0). Indeed, the equilibrium concentration of strategy R0 for combination D is $p_{\rm R0}\approx 0.018$.

\subsection{Finite populations}

\begin{figure*}[htbp] 
   \centering
   \includegraphics[width=\linewidth]{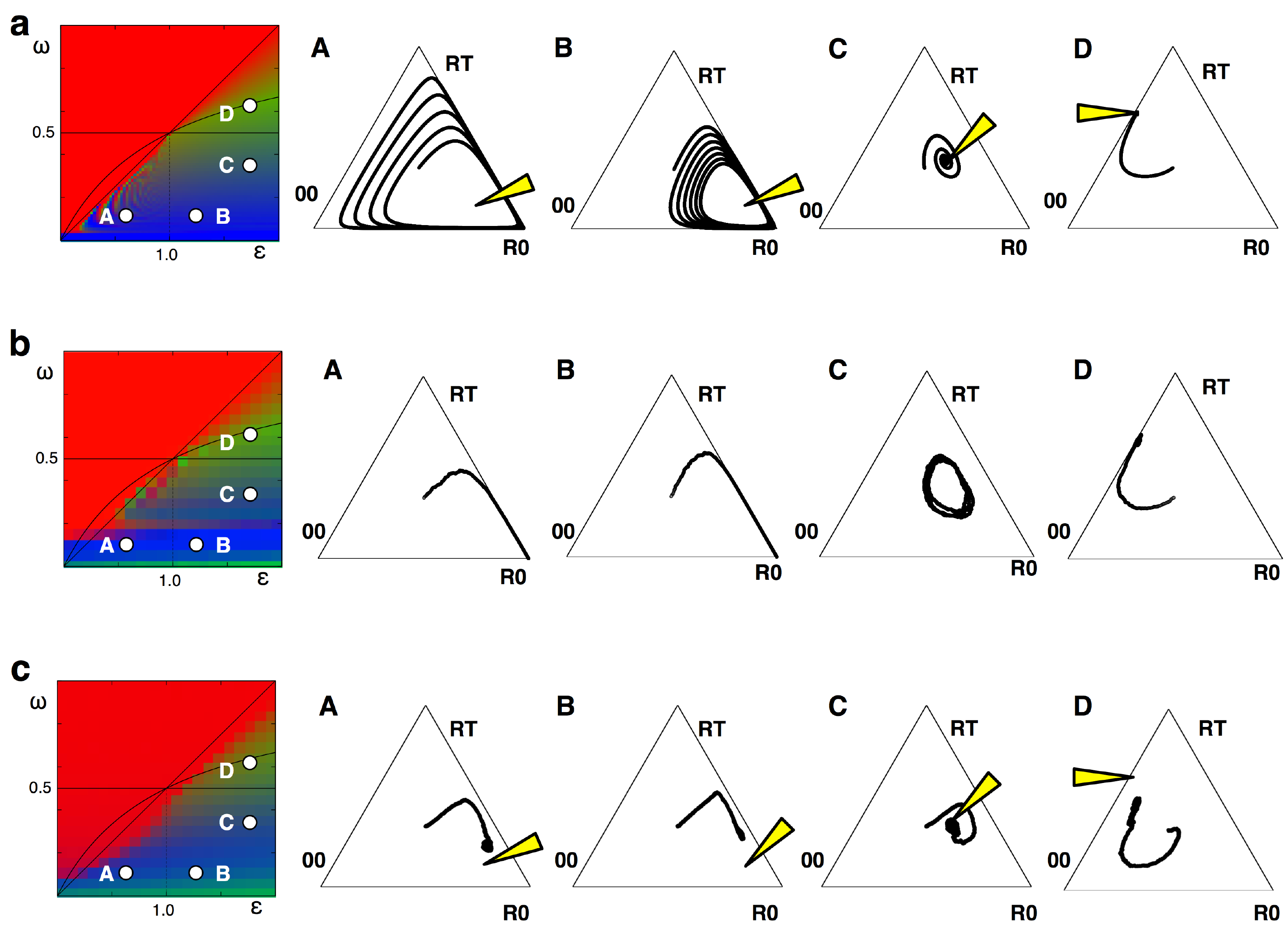} 
   \caption{(a) Left: Dominating strategy as determined by replicator equations as a function of cost $\omega$ and benefit $\veps$, where the density of each pure strategy is indicated by the intensity of the three colors red (00), green (RT), and blue (R0), given the starting condition where each strategy has 1/3 of the population. Regions as in Fig.~\ref{zeeman}.  Right: population fraction trajectories obtained from iterating the replicator equations for four parameter combinations $(\veps,\omega)$: $A=(0.75,0.125), B=(1.25,0.125), C=(1.75,0.375)$ and $D=(1.75,0.625)$, whose locations are shown on the left. Arrows indicate the location of the fixed points. (b) Dominating strategies in a finite population-size agent-based simulation of competition between pure strategies in a mell-mixed population (left, 20 runs of 5,000 updates per pair of parameters, for 21$\times 21$ parameter combinations.). Right: the trajectories for the four parameter combinations picked in (a) show stable coexistence for points C and D, but extinction of strategies for A and B. (c) Left: Darwinian evolution of probabilistic strategies in an agent-based simulation at fixed mutation rate (left, 20 replicates per pair of parameters, 21$\times 21$ parameter combinations), where colors indicate average fixed point probabilities in the scheme of (a). The trajectories (right) represent the average probabilities to engage in the plays 00, R0, and RT on an (averaged) line of descent, not trajectories of population frequencies. Arrows indicate the predicted fixed point. Average of 50 replicate trajectories for 100,000 updates for each of the example parameters.}
  \label{fig4}
\end{figure*}

If finite populations of pure strategy mixtures are simulated using agent-based methods as described earlier for the two-player game (population size 16,384), the dynamics are unchanged from the infinite population-size limit for the parameter region II. However, the heteroclinic orbits that we observed in region Ib collapse when populations are finite (as was noticed previously in Ref.~\cite{ClaussenTraulsen2008} for a generic RPS game in region Ib), because both 00 and RT go to extinction at the R0 fixed point (see Fig.~\ref{fig4}b) if the initial population consists of each strategy in equal concentration. Because of the nature of the flow, R0 is the surviving strategy for almost all initial conditions, reflecting a principle of ``survival-of-the-weakest" discussed recently in the context of cyclic stochastic games~\cite{Berretal2009}. Interestingly, as we approach the boundary of region IVb, fortunes are reversed: now R0 is dispensable and the game reverts to the two-player coordination game, where either RT or R0 survive, depending on the initial condition. 

Similar dynamics were also observed in experimental populations of sensitive, resistant, and toxin-producing bacteria engaged in an RPS game~\cite{Kerretal2002} as long as dispersal was high, which corresponds to the well-mixed case that we study here. However, loss of diversity (strategy extinction) can also occur in region Ia where the fixed point is stable, as long as the orbits have a high probability of touching the boundary before spiraling in. This dynamic is seen for point B in Fig.~\ref{fig4}b. For region III the finite population size does not alter the phase portrait appreciably: the fixed point in the interior is stable, and even when one of the equilibrium concentrations is small (as is the case for parameter combination D), the minority strategy does not disappear for the population size we studied. Of course, for sufficiently small populations,  R0 has a chance of extinction even in region IIIb.

\subsection{Stochastic strategies} \label{stoch}
As mentioned, cellular strategies in biology are often stochastic rather than pure. The decision to contribute to the stalk or the spore in the evolutionary game that {\it Dictyostelium} plays when resources become scarce is fully stochastic for example ($p\approx0.5$ in the wild type~\cite{Strassmannetal2000}).  This type of stochasticity is different from simulating errors of execution and perception~\cite{AxelrodDion1988}, that focus on small deviations from the deterministic scenario, and have been used extensively in the economical literature (see, e.g.,~\cite{HarsanyiSelten1988}). Stochastic strategies are often called ``mixed" strategies in the literature, but we prefer not to use this nomenclature because ``mixed" often evokes the idea of a mix of pure strategies (a polymorphic population). The fixed points of populations that use stochastic strategic have been studied less extensively than the phase portraits of deterministic populations, but some important results are known~\cite{Hines1980,Zeeman1981,Thomas1985,Bomze1990,Cressman1990}. For example, define a stochastic strategy ${\mathbf S}(q)=(q_{00},q_{\rm R0},q_{\rm RT})$ where the $q_i$ are the probabilities for this individual to engage in the plays $i$,  and let  its frequency in the population be $f(q)$. Then the population mean strategy is
\be
\bar {\mathbf S}=\sum_q f(q) {\mathbf S}(q)\;,
\ee
where we have assumed a discretization of strategy space (these averages can be generalized to continuous strategy spaces, see, e.g.~\cite{Hines1980,Zeeman1981}). 

If $\bfp$ is an ESS of the deterministic game defined by payoff matrix $E$, then the population mean (stochastic) strategy $\bar {\mathbf S}$ is a locally stable equilibrium of the mean strategy evolution, defined by~
\be
\frac{\dot f(q)(t)}{f(q)}=\left({\mathbf S}(q)-\bar {\mathbf S}\right)\cdot E \bar {\mathbf S} 
\ee
if and only if 
\be
\bar {\mathbf S}=\bfp\;.
\ee 
In other words, the fixed points of the deterministic game are the fixed points (in the sense that the population is ``fixed" at $\bar {\mathbf S}$) of the stochastic game~\cite{Zeeman1981}. As a consequence, if $\bfp$ is an ESS, then so is $\bar {\mathbf S}$. However, nothing is known (to our knowledge) about the stability of fixed points of the stochastic game that are not attractors for the deterministic game. At the same time, even when the population is fixed at $\bar {\mathbf S}$, the population composition can change neutrally as long as the average $\bar {\mathbf S}$ is preserved.

To simulate the Darwinian evolution of probabilistic strategies  we introduce separate loci for the `R' and `T' gene, encoding the probabilities $q_R$ and $q_T$. Thus, a `00' phenotype is expressed with probability $q_{00}=(1-q_R)(1-q_T)$, while the R0 phenotype has $q_{\rm R0}=q_R(1-q_T)$ and so forth. While we therefore also simulate the possibility of a `0T' phenotype, it never plays a dominant role in the dynamics as expected. Like in our simulation of the evolution of the two-player game, strategies play four random other players in the population to determine the fitness of the genotype $(q,q_T)$, which determines the probability with which this strategy leaves offspring in the next generation. The probabilities $q_R,q_T$ can be viewed as the stochastic decisions encoded by two different (and independent) genetic pathways with many genes, but rather than simulating how the genetic networks encode these probabilities, we evolve them directly. 

As before, we evolve the strategies using a death-birth Moran process, by removing at every update a random 2\% of a population of 1,024 genotypes in a well-mixed population, and replacing them with a set of genotypes that were picked (probabilistically, according to fitness)
from the set of players that survived the 2\% removal (we use a death-birth process rather than the more common birth-death process to avoid the awkward case that a genotype is born from a deleted type). When a genotype is not replaced, fitness accumulates from playing more games, but this does not result in a skewed age distribution as removal is random.

To determine the mean ``fixed point" strategy for each of 21$\times$ 21 parameter configurations (Fig.~\ref{fig4}c, leftmost panel), we first reconstruct the line of descent (LOD) of each of 20 replicate populations evolved for 500,000 updates, by picking the dominating strategy genotype and follow its ancestral line all the way back to the seed genotype (a strategy with $q_R=q_T=0.5$). The line of descent recapitulates the evolutionary history of that particular experiment, as it contains the sequence of mutations that gave rise to the successful strategy at the end of the run (see, e.g.,~\cite{Lenskietal2003}). The LOD also defines a trajectory in strategy space, but rather than being a smooth curve the LOD is jagged and jumps between probabilities. When averaging the LODs across runs to obtain the fixed point,  we first discard the last 50,000 updates for each (in order to make sure that our LOD has coalesced) and then average the genotypes of the LOD after the first 250,000 updates, in order to ensure that the trajectory settled on the fixed point (the procedure is described in more detail in~\cite{Iliopoulosetal2010}).

To obtain the strategy trajectory $\bar {\mathbf S}$ for the four parameter combinations A-D used earlier, we collect the plays of each agent in the population at each update (instead of collecting the frequencies $f(q)$ of each strategy  ${\mathbf S}(q)$), as the plays faithfully recapitulate the genotype in a well-mixed population~\cite{Iliopoulosetal2010}. The mean population strategy is plotted for the first 100,000 updates of an average of 50 replicate experiments in Fig.~\ref{fig4}c (right four panels).   As in previous work studying the evolution of stochastic strategies in the iterated Prisoner's Dilemma~\cite{Iliopoulosetal2010}, the population strategy moves to an evolutionary fixed point. In the present case (a non-iterated game),  the fixed point is at (or close) to the Nash equilibrium of the deterministic game. While the fixed points B-D are attractors (but not ESSs), fixed point A is repulsive. For the stochastic game, however, this fixed point turns out to be an attractor. 

In hindsight, this is not surprising when the competition is viewed in terms of an adaptive dynamics formalism (see, e.g.,~\cite{Geritzetal1998,DieckmannMetz2006}) as the population does not engage in a competition between three strategies that exclude each other. Rather, the population is fairly monomorphic, centered around the single stochastic strategy with probabilities given by the Nash fixed point. Thus, rather than witnessing positive frequency-dependent selection in a competition of three types, we see stable mutation-selection balance of a single type.   

We note that the diversity of plays that we observe appears to contradict in part previous conclusions that local dispersal promotes diversity in microbial dynamics of the RPS-type~\cite{Kerretal2002}. These authors concluded that spatial interaction environments (nearest-neighbor as opposed to well-mixed) promote diversity, observing coexistence when a well-mixed population simulation predicted extinction of two out of the three strategies, for parameter combinations that we estimate puts their population in the region IB or IVb, with a repelling fixed point. Yet, when phenotypes are expressed probabilistically, strategies are stabilized. However, while we know that at least the decision to express the `T' type is stochastic in microbial colicin phenotypes,  the expression of the `R' type may be deterministic. Thus, our simulations cannot be directly compared to the experiments of Ref.~\cite{Kerretal2002}. We should also keep in mind that we have not explicitly taken into account the effect of spatial structure~\cite{Iwasaetal1998} here, but instead simply varied the payoff $\veps$. However, the loss of diversity was predicted to occur in the well-mixed mode, which we show to be stable instead when decision are probabilistic. 

\section*{Conclusions}
We have studied the fixed points and phase portraits of populations engaged in game-theoretical dynamics inspired by the microbial ``suicide-bomber" game. Depending on the physical parameter values that determine the payoffs of different decisions,  all the well-known games of the two-player scenario are covered. When extending to a three-player game by decoupling resistance from toxin production, we study the fixed points and stability not of all possible three-player games, but a subset of seven of 38 possible phase portraits. When strategies are allowed to be stochastic rather than deterministic, we observe that the population mean strategy moves toward the deterministic strategy fixed point, but repelling fixed points become attractive when decisions are probabilistic. We stress again that the original ESS concept is lacking for the three-player game, as none of the fixed points (even the attractive ones) are ESSs in the sense of Maynard Smith~\cite{MaynardSmith1982}.

It is difficult to ascertain which dynamics we should expect in natural populations, as the cost of production of the colicin as well as the cost of resistance vary considerably~\cite{FeldgardenRiley1999,Nahumetal2011}.  In general, we should expect that the costs are different, and that the efficiency of killing due to bacteriocins is not 100\%, so that the more general model (\ref{Mat3}) should be used. But given these caveats, our predictions of evolutionary fixed points as a function of cost and benefit parameters should be testable by dedicated experiments of the sort described in~\cite{Kerretal2002}, by changing the parameters of the evolutionary game (for example by experimentally varying the cost of resistance or benefit of the toxin). Recently, Nahum et al.~\cite{Nahumetal2011} have studied the evolution by natural selection of the parameters that define the game, and found that the resistant type (here, type R0)  tended to evolve towards more restrained interactions (higher $\delta$) compared to populations that evolved in the absence of interactions, at least in the case that migration was restricted so that spatial effects are present. This appears to be a direct verification of the ``survival-of-the-weakest" principle~\cite{Freanetal2001,Berretal2009} that is applicable in environments where finite population size introduces stochasticity, in parameter region Ib of the three-player game. Clearly it will be interesting to see more general evolutionary trajectories within the $\omega,\veps$-space (or the more general $\delta,\omega,\veps$-space), under different realistic constraints for spatial interactions, as well as costs of resistance and toxin production that are constrained to lie within biologically reasonable assumptions. It has previously been shown that changing environmental conditions can change the dynamics fro a prisoner's Dilemma to a snowdrift game (moving from region I into region III in Fig.~\ref{2dyn}) in the two-player game~\cite{Goreetal2009,WangGoldenfeld2011}. It would be interesting to test whether populations can be coaxed to move from one Zeeman class to another in the three-player game, simply be changing the selective pressures acting on the system.

\section*{Acknowledgements} We thank Benjamin Kerr for extensive discussions. This work was supported in part by the National Science Foundation's Frontier in Integrative Biological Research grant  FIBR-0527023 and NSF's BEACON Center for the Study of Evolution in Action, under contract No. DBI-0939454. We wish to acknowledge the support of the Michigan State University High Performance Computing Center and the Institute for Cyber Enabled Research.

\bibliography{PD}
\end{document}